\documentclass[submission, Proceedings]{SciPost}

\hypersetup{
    colorlinks,
    linkcolor={red!50!black},
    citecolor={blue!50!black},
    urlcolor={blue!80!black}
}

\usepackage[bitstream-charter]{mathdesign}
\DeclareSymbolFont{usualmathcal}{OMS}{cmsy}{m}{n}
\DeclareSymbolFontAlphabet{\mathcal}{usualmathcal}

\newcommand{\ev}{\mbox{eV\,}$c^{-2}$}
\newcommand{\eve}{\mbox{eV$_{\rm ee}$}}
\newcommand{\um}{\mbox{$\mu$m}}

\begin{document}

\begin{center}{\Large \textbf{
Even Lighter Particle Dark Matter\\
}}\end{center}

\begin{center}
Alvaro E. Chavarria\textsuperscript{1$\star$}
\end{center}

\begin{center}
{\bf 1} Center for Experimental Nuclear Physics and Astrophysics, University of Washington, Seattle, United States
\\
* chavarri@uw.edu
\end{center}

\begin{center}
\today
\end{center}


\definecolor{palegray}{gray}{0.95}
\begin{center}
\colorbox{palegray}{
  \begin{tabular}{rr}
  \begin{minipage}{0.1\textwidth}
    \includegraphics[width=30mm]{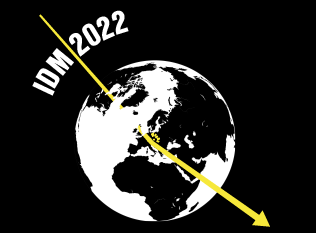}
  \end{minipage}
  &
  \begin{minipage}{0.85\textwidth}
    \begin{center}
    {\it 14th International Conference on Identification of Dark Matter}\\
    {\it Vienna, Austria, 18-22 July 2022} \\
    \doi{10.21468/SciPostPhysProc.?}\\
    \end{center}
  \end{minipage}
\end{tabular}
}
\end{center}

\section*{Abstract}
{\bf
We report on recent progress in the search for dark matter particles with masses from 1\,M\ev\ to 1\,G\ev .
Several dark matter candidates in this mass range are expected to generate measurable electronic-recoil signals in direct-detection experiments.
We focus on dark matter particles scattering with electrons in semiconductor detectors since they have fundamentally the highest sensitivity due to their low ionization threshold.
Charge-coupled device (CCD) silicon detectors are the leading technology, with significant progress expected in the coming years.
We present the status of the CCD program and briefly report on other efforts.
}

\section{Introduction}
\label{sec:intro}

Weakly interacting massive particles (WIMPs)~\cite{Kolb:1990vq, Griest:2000kj, Zurek:2013wia} with masses $m_\chi>1$\,G\ev\ have been the leading dark matter (DM) candidate for many decades.
Direct searches for their scattering with atomic nuclei in underground detectors have progressed significantly throughout this time, without any discovery.
Although the WIMP direct detection program continues unabated~\cite{Akerib:2022ort}, a fraction of the scientific community has recently turned its attention to the possibility that DM particles are significantly lower in mass than previously considered~\cite{Essig:2022dfa}.
Theoretical studies have uncovered several well-motivated scenarios where DM particles with $m_\chi<1$\,G\ev\ can be created in the early universe and detected today in the laboratory.
This has spurred a number of experimental searches with existing detector technologies, and motivated the development of novel technologies with improved sensitivity to the low-energy depositions from light DM particles.
The leading new technology of skipper charge-coupled devices (CCDs) is reaching maturity with the deployment of the first sizable detectors underground, which are expected to improve in sensitivity to light DM particles by orders of magnitude in the coming years.

\section{Direct detection signals from low-mass DM particles}
\label{sec:signals}

For DM with $m\chi<1$\,G\ev , the traditional approach to search for the energy deposited in the target by nuclear recoils from DM scattering becomes limited because the mismatch between the mass of the DM particle and the nucleus does not allow for efficient transfer of energy.
Recently, other direct-detection mechanisms have been proposed to extend sensitivity to lower DM masses.
The first is to search for 
electrons or photons emitted in the DM-nucleus interaction, which can carry a significant fraction of the DM particle's kinetic energy and appear above the detection threshold.
The ``Migdal effect''~\cite{Ibe:2017yqa} is the process where an electron from the recoiling atom is emitted, with a probability in silicon of $\mathcal{O}(10^{-5})$~\cite{Knapen:2020aky}.
The probability of photon emission (the ``Bremsstrahlung'' process) is significantly smaller~\cite{Kouvaris:2016afs}.
Another possibility is to look directly for the scattering of low-mass DM particles with electrons in the target~\cite{Essig:2017kqs, Essig:2015cda}.
Atomic electrons are lighter than nuclei and, since they have a momentum distribution, there are regions of phase space where the electron can take a significant fraction of the DM particle's kinetic energy.
DM-e interactions naturally arise in hidden-sector vector-portal DM theories~\cite{Chu:2011be}, where the DM-e scattering is mediated by a ``hidden photon.''
The hidden photon itself could also constitute the DM and be absorbed by atomic electrons~\cite{An:2014twa, Hochberg:2016sqx, Bloch:2016sjj}.
Since all these DM signals are electronic recoils in the target, several experiments have placed exclusion limits on sub-GeV DM from their measured electronic-recoil spectra~\cite{DarkSide:2018ppu, DAMIC:2019dcn, XENON:2019gfn, SENSEI:2020dpa}.

\begin{figure}[t]
\centering
\includegraphics[width=\textwidth]{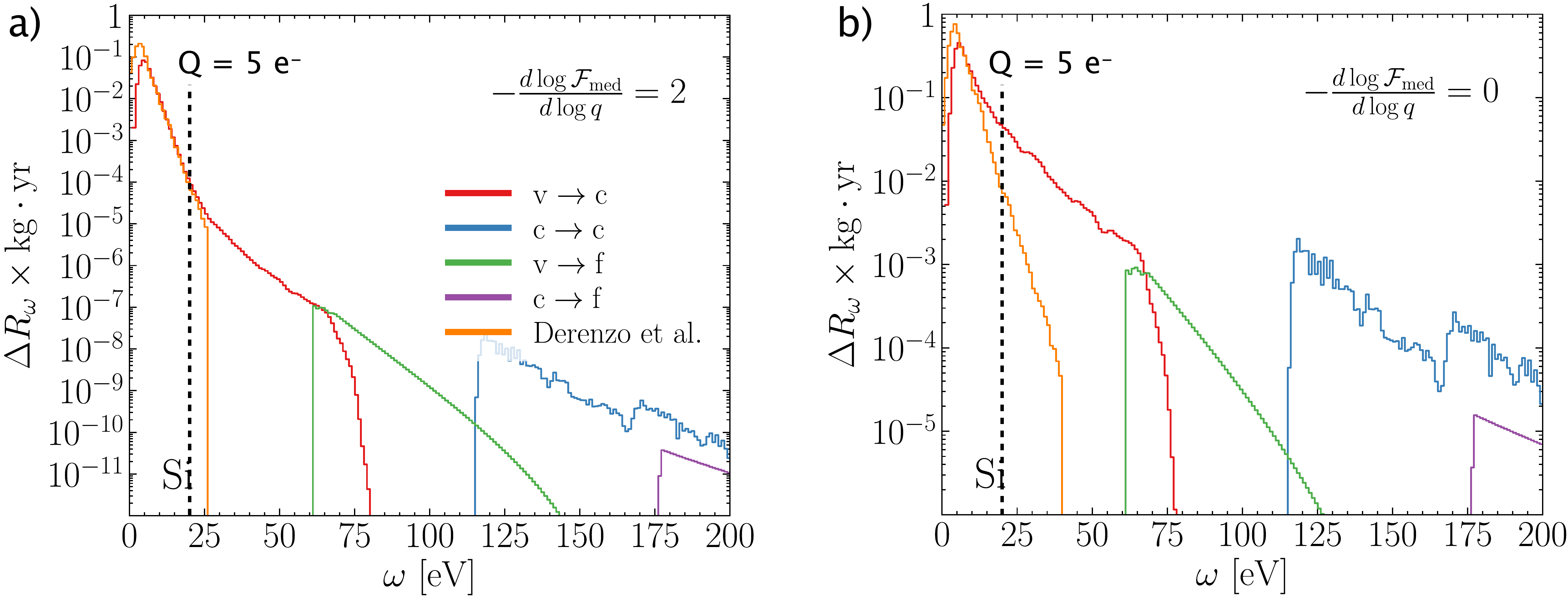}
\caption{Deposited-energy ($\omega$) spectra by a 1\,G\ev\ DM particle scattering with electrons in silicon via an ultra-light ({\bf a}) or heavy ({\bf b}) mediator. An ionization signal of $Q=5$\,$e^-$ corresponds to $\omega \sim 20$\,eV. Calculated with the {\tt EXCEED-DM} code. Figures adapted from Ref.~\cite{Griffin:2021znd}.}
\label{fig:spectra}
\end{figure}

The kinetic energy of a DM particle in the galactic halo is $E_K\sim10^{-6} m_\chi c^2$.
Most often, only a fraction of $E_K$ is transferred to the electron, with the predicted deposited-energy spectrum highly peaked at the lowest energies.
Fig.~\ref{fig:spectra} shows the spectrum from a 1\,G\ev\ DM particle scattering with electrons in silicon predicted by the {\tt EXCEED-DM} code~\cite{Griffin:2021znd}.
Most of the spectrum\textemdash even at 1\,G\ev \textemdash is below 10\,\eve , which corresponds to only a few ionized charges in silicon.
Thus, detectors capable of counting single ionized electrons (already achieved in semiconductor and noble-element direct-detection technologies) are needed to perform sensitive searches for low-mass DM-e interactions.
Semiconductor detectors have a fundamental advantage since the ionization threshold is $\sim$1\,eV compared to $>$10\,eV for noble elements.
Furthermore, the momentum distribution of the electrons in the semiconductor valence band are better ``kinematically matched'' to the DM flux~\cite{Essig:2015cda}.
For this reason, the current sensitivity of silicon CCD detectors to DM-e scattering surpasses that of noble liquids for most of the parameter space (specially for smaller DM masses and lighter mediators) even though their target exposures are orders of magnitude smaller ($\mathcal{O}$(g$\cdot$day) vs. $\mathcal{O}$(tonne$\cdot$day)).

\section{CCDs to search for DM}
\label{sec:ccds}

\begin{figure}[t]
\centering
\includegraphics[width=\textwidth]{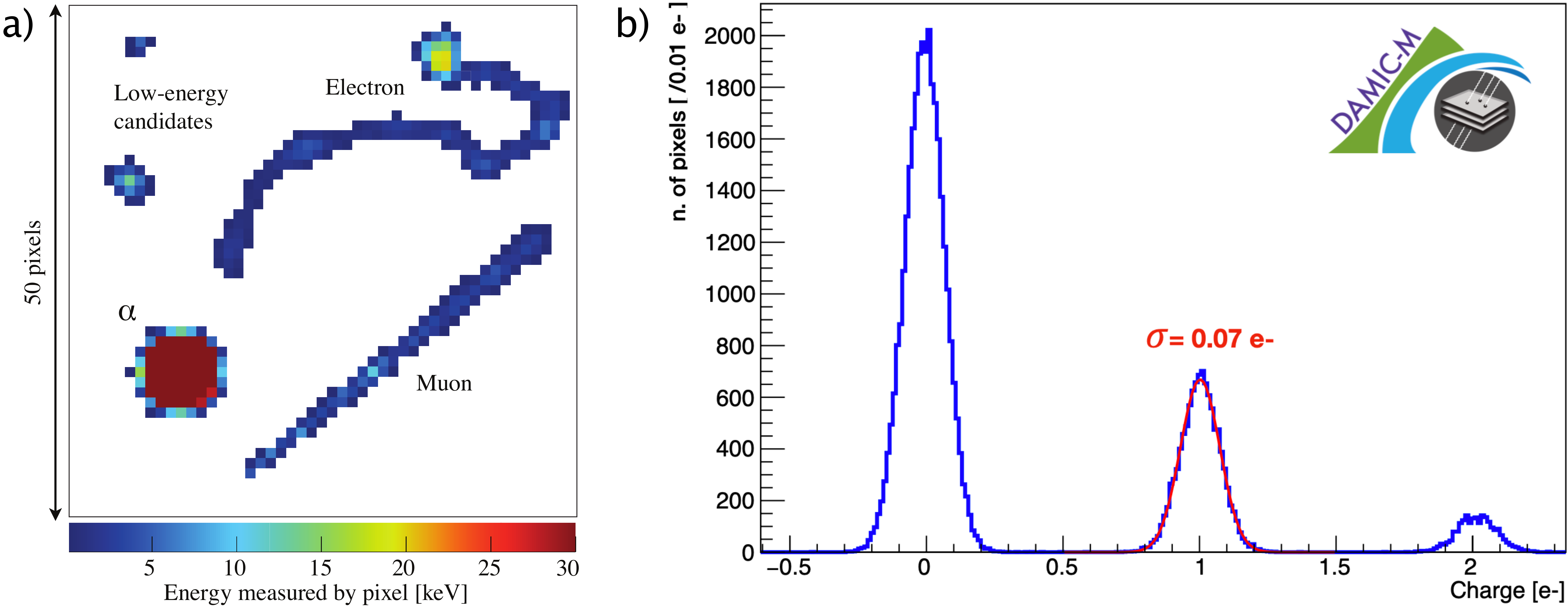}
\caption{{\bf a)}~A 50$\times$50 pixels segment of a CCD image from an exposure in the surface lab. Different types of particles (labeled) can be distinguished by their topology. The orientation of the muon track in $z$ can be reconstructed from the fact that the track is less (more) diffuse through the point where it crosses the front (back) surface of the CCD. {\bf b)}~Pixel-value distribution of a DAMIC-M skipper CCD, which demonstrates the capability to count charges per pixel with high resolution.}
\label{fig:ccd}
\end{figure}

Charge-coupled devices (CCDs) are monolithic solid-state silicon imaging devices that measure the free charges generated in their fully-depleted active target.
Particles generate free charges (e-h pairs) in the CCD active region by ionization, with a minimum energy to ionize an e-h pair of 1.2 eV and one e-h pair generated on average for every 3.8\,eV of kinetic energy deposited by a recoiling electron. 
The free charges are then drifted by an electric field toward the pixel array.
Since charge diffuses laterally with time as it drifts, the spread of the charge cluster in the image ($x$-$y$ plane) is positively correlated to the depth ($z$) of the interaction.
Thus, CCDs can, in principle, provide the deposited energy and $(x, y, z)$ location of particle interactions in the bulk silicon.
In reality, the spatial resolution may be limited in one or more axes depending on the readout mode and event energy.
Fig.~\ref{fig:ccd}a shows sample particles tracks from an image exposure in the surface laboratory.

All CCDs currently used for DM searches were developed by Berkeley Lab's Microsystems Laboratory and fabricated by Teledyne DALSA.
The devices feature a rectangular array of pixels, each of size 15$\times$15\,\um$^{2}$, with a total area of $\mathcal{O}$(10\,cm$^2$) and a fully-depleted active region of 675\,\um.
Other details of the CCD design and fabrication process can be found in Ref.~\cite{1185186}.

CCDs are the most promising technology in the search for dark-sector DM from their interactions with electrons for two main reasons.
The first is the extremely low readout noise ($\sim$0.05\,$e^-$) of ``skipper'' CCDs, first demonstrated by the SENSEI Collaboration~\cite{Tiffenberg:2017aac}, which allows the devices to count with high resolution the number of charges collected by every pixel (Fig.~\ref{fig:ccd}b).
The second was the demonstration by the DAMIC Collaboration~\cite{DAMIC:2016lrs,DAMIC:2016qck} that sizable exposures ($\sim$10\,kg$\cdot$day) with extremely low backgrounds from leakage current (few $e^-$ per mm$^2$$\cdot$day~\cite{DAMIC:2019dcn}) and ionizing radiation (few events per k\eve $\cdot$kg$\cdot$day~\cite{DAMIC:2021crr}) can be acquired with a CCD array shielded deep underground.
The combination of the unprecedentedly low ionization threshold with low background in sizable exposures provides CCD detectors with the current best sensitivity to MeV-scale DM particles.

Dark matter searches are performed by comparing the charge (energy) distributions of individual pixels or pixel clusters against a background model that includes instrumental noise and ionizing backgrounds.
Ionizing backgrounds include charged particles from natural radioactivity that deposit their kinetic energy in the bulk silicon, and the photoelectric absorption of optical and near-infrared photons~\cite{Du:2020ldo}.
To construct the signal and background models for DM searches, detailed studies have been performed on the response of CCDs to ionization~\cite{Chavarria:2016xsi, Ramanathan:2020fwm, Rodrigues:2020xpt, DAMIC:2021crr, DAMIC-M:2022xtp}, and on instrumental effects that may generate charges in CCDs~\cite{SENSEI:2021hcn}.

\section{The DAMIC program and other CCD detectors}
\label{sec:damic}

DAMIC pioneered the search for DM with CCDs, deploying detectors at the SNOLAB underground laboratory since 2012, with the final installation of seven 16\,Mpix CCDs (40\,g silicon target) in 2017.
These detectors featured CCDs with conventional readout, which achieved a pixel noise of 1.6\,$e^-$ R.M.S.
Details of the DAMIC setup at SNOLAB are presented in Refs.~\cite{DAMIC:2016lrs, DAMIC:2021crr}.
DAMIC made steady progress in the reduction and understanding of ionizing backgrounds in the detector, including an extensive radioassay program of all detector components~\cite{DAMIC:2021crr}, a measurement of the cosmogenic activation of silicon~\cite{Saldanha:2020ubf}, and the development of data analysis techniques to identify sources of background.
Of particular relevance are the measurements of radiocontaminants in the CCDs, \emph{e.g.}, surface/bulk $^{210}$Pb and bulk $^{32}$Si, by searching for spatio-temporal correlations between decays~\cite{DAMIC:2020wkw}.
The identification of these isotopes\textemdash with characteristic time between decays of $\sim$10\,days\textemdash is only possible because of the high spatial resolution and solid-state target of CCDs.
DAMIC's efforts in the mitigation of background resulted in a total (bulk) background in the final detector of 10 (5) events per k\eve $\cdot$kg$\cdot$day.
The understanding of backgrounds culminated with the construction of the first complete radioactive background model for a DM search with a CCD detector~\cite{DAMIC:2021crr}.

DAMIC performed in an 11\,kg$\cdot$day exposure the most sensitive direct search for weakly interacting massive particles (WIMPs) with masses in the range 1--9\,G\ev\ scattering with silicon nuclei\cite{DAMIC:2020cut}.
The background model describes the data remarkably well above 200\,\eve\ but there is a statistically significant (3.7\,$\sigma$) excess of events at lower energies, which is well described by a spatially uniform population of events with a rate of a few per kg$\cdot$day and an exponential spectrum down to the analysis threshold of  50\,\eve .
The measured spectrum\textemdash with calibrated nuclear-recoil energy scale~\cite{Chavarria:2016xsi}\textemdash is inconsistent with the standard WIMP interpretation of a previous excess of nuclear recoils reported by the CDMS-II Si experiment~\cite{CDMS:2013juh}, and directly constrains other DM interpretations.

In the search for hidden-sector DM particles interacting with electrons, DAMIC and SENSEI have been leapfrogging past each other in the last five years.
In 2017, DAMIC was the first experiment to search for DM interactions that produce as little as a one e-h pair in silicon, resulting in the first exclusion limit on the absorption of hidden photons with masses as small as 1.2\,\ev~\cite{DAMIC:2016qck}.
SENSEI released the first results on DM-e scattering with data from a 0.1\,g prototype skipper CCD operating on the surface at Fermilab in 2018~\cite{Crisler:2018gci}, constraining the existence of DM particles with masses as small as 0.5\,M\ev .
The results were improved in 2019 by operating the device with significantly reduced background in the shallow underground site of the MINOS cavern at Fermilab~\cite{SENSEI:2019ibb}.
DAMIC followed with world-leading exclusion limits on DM-e scattering later in 2019~\cite{DAMIC:2019dcn}, made possible by leveraging its lower leakage current and larger exposure to make up for the higher noise of CCDs with conventional readout.
These results were greatly surpassed in 2020 by SENSEI after operating a 2\,g scientific skipper CCD (with comparable leakage current to DAMIC) in the MINOS  cavern, finally demonstrating the potential of skipper technology~\cite{SENSEI:2020dpa}.

\begin{figure}[t]
\centering
\includegraphics[width=\textwidth]{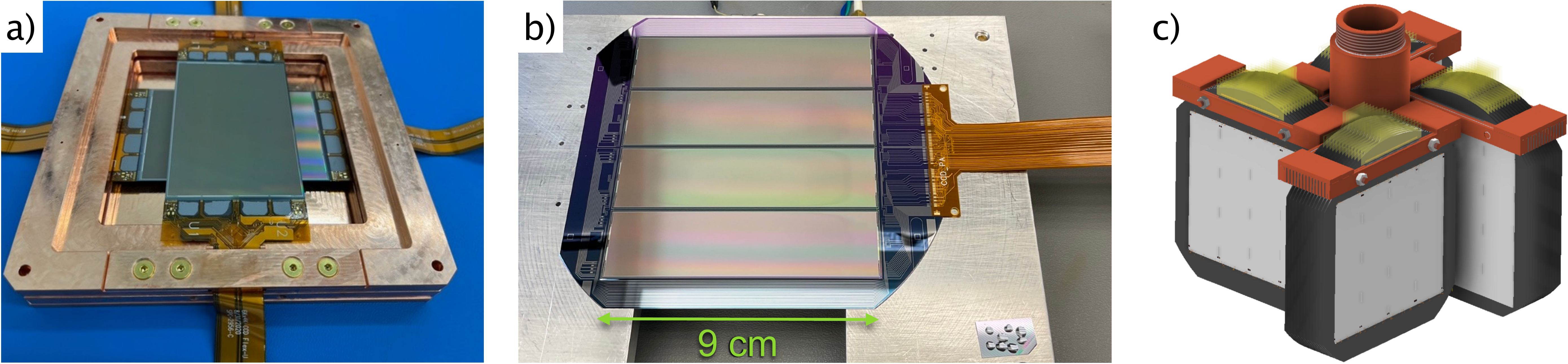}
\caption{{\bf a)}~Two 24\,Mpix CCDs installed in their copper box before deployment in DAMIC-M's Low Background Chamber (LBC). {\bf b)}~Prototype DAMIC-M module with four preproduction CCDs on a pitch adapter at the University of Washington. {\bf c)}~Proposed arrangement of the 52 CCD modules of the final DAMIC-M detector.}
\label{fig:damicm}
\end{figure}

\begin{figure}[t]
\centering
\includegraphics[width=\textwidth]{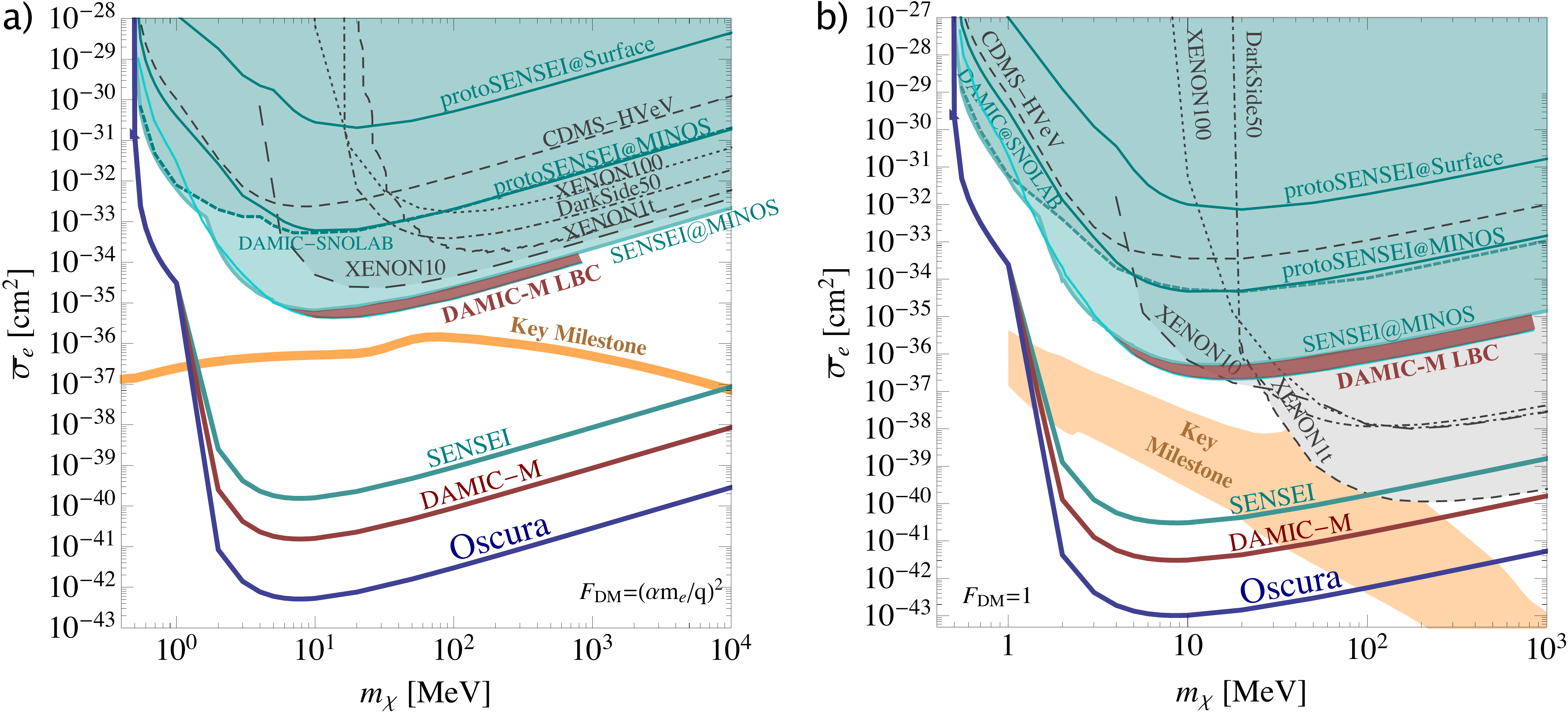}
\caption{ Exclusion limits on the DM-e scattering cross section from existing and planned experiments. The parameter space for scattering via an ultra-light ({\bf a}) or heavy ({\bf b}) mediator are presented. Teal shaded regions correspond to published results, while the red shaded region is the new parameter space excluded by the preliminary DAMIC-M result from IDM 2022. The solid lines are forecasts for SENSEI, DAMIC-M and Oscura. Figure adapted from Ref.~\cite{Aguilar-Arevalo:2022kqd}.}
\label{fig:limits}
\end{figure}

Throughout 2021, skipper-CCD detectors with significant target mass ($\sim$20\,g) were deployed deep underground by the DAMIC-M and SENSEI Collaborations.
DAMIC-M, the successor of DAMIC, deployed its first skipper CCD detector in the Modane Underground Laboratory (LSM) in France.
Fig.~\ref{fig:damicm}a shows the two 24\,Mpix large-format skipper CCDs (9\,g each) in the Low Background Chamber (LBC).
This setup produced the preliminary DAMIC-M exclusion limits for DM-e scattering shown in Fig.~\ref{fig:limits}, which were presented by D.~Norcini at IDM 2022 (this conference).
The LBC will also serve as a testbed to decrease ionizing backgrounds from natural radioactivity from the current level of 10 events per k\eve $\cdot$kg$\cdot$day to DAMIC-M's goal of 0.1 events per k\eve $\cdot$kg$\cdot$day.
SENSEI installed 10 smaller skipper CCDs ($\sim$25\,g) in a new detector at SNOLAB, with a first science run starting in summer 2022.
SENSEI's final goal is a 100\,g target with a background of 5 events per k\eve $\cdot$kg$\cdot$day to realize the forecast in Fig.~\ref{fig:limits}.
Two more DAMIC-M large-format skipper CCDs were deployed in the existing DAMIC cryostat at SNOLAB with the specific task of performing a more precise measurement that could shed light on the origin of the event excess reported in Refs.~\cite{DAMIC:2020cut, DAMIC:2021crr}.
Skipper CCDs will allow for a spectral measurement with a lower threshold of 15\,\eve , and higher resolution in $z$ for better discrimination between bulk and surface events.
The science run started in March 2022, with first results expected in early 2023.
This is a collaborative effort between DAMIC-M and SENSEI since, if unaddressed, this background will dominate the low-energy spectrum of future CCD DM detectors.

The ultimate goal of the DAMIC-M program is the deployment of a large array of 52 skipper-CCD modules at LSM to accrue kg$\cdot$year exposures with a 2 or 3 $e^-$ threshold.
Fig.~\ref{fig:damicm}b shows a prototype DAMIC-M module with four 9\,Mpix CCDs (13.5\,g of silicon per module), while Fig.~\ref{fig:damicm}c shows a preliminary design of the arrangement of the 52 modules in the cryostat.
Improvements in detector design and construction are needed to meet the stringent background requirements of the detector.
These include: i)~development and fabrication of low-radioactivity flex cables, ii)~the use of ultra-pure copper electroformed underground for the CCD holders, iii)~the use of shielded containers throughout CCD fabrication to mitigate cosmogenic activation, and iv)~the assembly of the CCD  modules underground in a radon-free environment to mitigate surface contamination.
Two preproduction CCD runs to confirm the quality of the devices were already completed, with final production of DAMIC-M CCDs starting in late 2022.
The detector design is in its final stages with detector construction expected throughout 2023 and detector commissioning in 2024.

Beyond SENSEI and DAMIC-M, there are plans for a larger Oscura detector at SNOLAB, capable of acquiring a 30\,kg$\cdot$year exposure with a 2 or 3 $e^-$ threshold~\cite{Aguilar-Arevalo:2022kqd}. 
Current R\&D activities focus on scaling up the existing technology, with steadfast progress in the electronics required for multiplexing and processing the signals from the detector's 24,000 skipper amplifiers.
Fig.~\ref{fig:limits} shows forecasts for the sensitivity of Oscura to DM-e scattering, together with DAMIC-M and SENSEI.

\section{Other detector technologies}
\label{sec:other}

Cryogenic calorimeters operating at mK temperatures with very low noise phonon sensors have demonstrated low ionization thresholds by operating in ``high-voltage'' mode, where the phonon signal is proportional to the number of ionized charges because of the Neganov-Trofimov-Luke (NTL) effect~\cite{SuperCDMS:2013eoh}.
With this strategy, a 1\,g silicon ``HVeV'' detector demonstrated 0.03\,$e^-$ R.M.S. noise in the ionization signal, with the capability of counting single charges with high resolution~\cite{Ren:2020gaq}, like a skipper CCD.
The SuperCDMS Collaboration published exclusion limits on DM-e scattering from two runs of HVeV detectors that acquired $\sim$g$\cdot$day exposures in surface laboratories~\cite{SuperCDMS:2018mne, SuperCDMS:2020ymb}.
The EDELWEISS Collaboration operated a larger 30\,g germanium detector in HV mode with 0.53\,$e^-$ R.M.S. noise on the ionization signal, resulting in the best exclusion limit on DM-e scattering with a germanium target and with a cryogenic calorimeter~\cite{EDELWEISS:2020fxc}.
This was possible despite the higher noise because the larger detector i)~allowed for a larger exposure, and ii)~had a lower surface-background rate due to its smaller surface-to-volume ratio.
The detector was also operated deep underground in LSM, which contributed to overall lower backgrounds.

The results from SuperCDMS in 2018~\cite{SuperCDMS:2018mne} and 2020~\cite{SuperCDMS:2020ymb}, and EDELWEISS in 2020~\cite{EDELWEISS:2020fxc} were competitive with CCD results published at the time~\cite{SENSEI:2019ibb, DAMIC:2019dcn}, but were significantly surpassed by the most recent results from SENSEI~\cite{SENSEI:2020dpa} and DAMIC-M (presented at IDM 2022).
Several challenges remain specific to cryogenic calorimeters.
Unlike CCDs, were the readout noise is decoupled from the overall size of the device, sufficiently low phonon noise for single-charge resolution in a detector of mass $\mathcal{O}$(10\,g) remains to be demonstrated.
Furthermore, phonon sensors are sensitive to ``heat-only'' events, which constitute a rising spectrum toward low energies.
The origin of these events was the focus of the recent series of EXCESS Workshops~\cite{Fuss:2022fxe}, and likely explanations include phonon bursts induced by stresses in the detector components~\cite{Anthony-Petersen:2022ujw}.
Nevertheless, there is active R\&D in the further development of these detectors, with future deployments planned for DM searches by EDELWEISS in LSM and SuperCDMS in the NEXUS shallow underground facility at Fermilab, and later in SNOLAB~\cite{SuperCDMS:2022kse}.

\section{Conclusion}
\label{sec:conclusion}

The search for low-energy ionization signals in direct-detection experiments has demonstrated to be a highly sensitive probe for the existence of sub-GeV DM particles.
Semiconductor detectors, which have the lowest ionization thresholds, have progressed dramatically in recent years.
In particular, silicon CCD detectors have moved past the R\&D stage with sizable detectors deployed deep underground in 2021.
A staged program aims to scale up in target size by many orders of magnitude in the next decade.
The CCD program is poised to uncover large regions of parameter space for DM particles with masses between 1\,M\ev\ and 1\,G\ev\ from their scattering with electrons, with significant potential for discovery.

\section*{Acknowledgements}
These proceedings report on the work of the DAMIC, DAMIC-M, SENSEI, Oscura, SuperCDMS and EDELWEISS Collaborations.
The author's DM research is supported by the United States National Science Foundation (NSF) through grant PHY-2110585 and the Department of Energy's Office of Science through the Dark Matter New Initiatives program.

\bibliography{myrefs.bib}

\nolinenumbers

\end{document}